# Broadband and fabrication-tolerant 3-dB couplers with topological valley edge modes

Guo-Jing Tang[1,†], Xiao-Dong Chen[1,†], Lu Sun[2,†], Chao-Heng Guo[1], Meng-Yu Li[1], Zhong-Tao Tian[1], Hou-Hong Chen[1], Hong-Wei Wang[2], Qi-Yao Sun[2], Ying-Di Pan[2], Xin-Tao He[1,*], Yi-Kai Su[2,*] and Jian-Wen Dong[1,*]

1, School of Physics & State Key Laboratory of Optoelectronic Materials and Technologies, Sun Yat-sen University, Guangzhou 510275, China.

2, State Key Laboratory of Advanced Optical Communication Systems and Networks, Department of Electronic Engineering, Shanghai Jiao Tong University, Shanghai 200240, China.

†These authors contributed equally to this work

*Corresponding author: hext9@mail.sysu.edu.cn, yikaisu@stju.edu.cn, dongjwen@mail.sysu.edu.cn

## ABSTRACT

3-dB couplers, which are commonly used in photonic integrated circuits for on-chip information processing, precision measurement, and quantum computing, face challenges in achieving robust performance due to their limited 3-dB bandwidths and sensitivity to fabrication errors. To address this, we introduce topological physics to nanophotonics, developing a framework for topological 3-dB couplers. These couplers exhibit broad working wavelength range and robustness against fabrication dimensional errors. By leveraging valley-Hall topology and mirror symmetry, the photonic-crystal-slab couplers achieve ideal 3-dB splitting characterized by a wavelength-insensitive scattering matrix. Tolerance analysis confirms the superiority on broad bandwidth of 48 nm and robust splitting against dimensional errors of 20 nm. We further propose a topological interferometer for on-chip distance measurement, which also exhibits robustness against dimensional errors. This extension of topological principles to the fields of interferometers, may open up new possibilities for constructing robust wavelength division multiplexing, temperature-drift-insensitive sensing, and optical coherence tomography applications.

# Introduction

Photonic integrated circuits provide an efficient platform for optical interconnection [1], optical sensing [2], integrated quantum circuits [3] and optical neural networks [4], and find broad applications in high-speed communication, internet of things, quantum information technology and artificial intelligence. Among the fundamental components in these circuits is the 3-dB coupler, which facilitates the equal splitting of signal power for on-chip information processing. Conventional 3-dB couplers, such as directional couplers (DCs), face challenges of limited operation bandwidth and weak robustness because the power partition is governed by the interference of two modes in coupled waveguides [5, 6]. Consequently, the splitting ratio has serious sensitivity to working wavelength and fabrication dimensional errors. To mitigate these issues, researchers have proposed and implemented various well-designed nano-structures [7-16], including adiabatic evolution structures, asymmetric waveguides, multiple sections and sub-wavelength structures. While the performance of the optimized 3-dB couplers is improved, the splitting ratios are still wavelength dependent due to the reliance of interference principles. To overcome such limitation, it is necessary to explore new physical principles beyond interference so as to achieve a broadband and fabrication-tolerant 3-dB coupler.

Topological nanophotonics, a recently flourishing research field that combines the topological physics with nanophotonics [17-22], offers exciting opportunities for realizing robust waveguides [23-32], on-chip microcavities [33-39] and advanced lasers [40-46]. As one of representative topological nanostructures, valley photonic crystals (VPCs) support edge modes with inter-valley-scattering suppression [47]. These valley edge modes are resilient against perturbations that do not mix the two valleys. It should be emphasized that VPCs cannot completely eliminate radiation losses resulting from arbitrary perturbations, which is predicted by an early theoretical study [17] and further confirmed through a recent experimental demonstration [48]. VPCs have found applications in various areas, with one notable example being the realization of broadband robust transport through sharp-bent waveguides, as well as the demonstration on the subsequent splitting phenomena [49-51]. Therefore, VPCs hold great promise for achieving wavelength-independent splitting, addressing the issues associated with operation bandwidth and fabrication tolerance.

In this work, we utilize VPCs to develop a topological 3-dB coupler. We thoroughly investigate the operational principles of such topological couplers and discover a scattering matrix insensitive to wavelength, resulting in the ideal 3-dB power splitting. This broadband and robust splitting ratio is achieved through the combination of valley-Hall topology and mirror symmetry. The 3-dB power splitting supports a wide wavelength range around 1550 nm and exhibits tolerance to dimensional errors, as compared to the case of conventional DCs. To further demonstrate its functionality, we propose an on-chip Michelson-like interferometer by utilizing the aforementioned topological coupler. We experimentally validate its capability of extracting length differences and demonstrate its robustness against dimensional errors. These findings highlight the resilience of the splitting ratio and the potential of realizing broadband and fabrication-tolerant topological 3-dB couplers in a wide variety of applications such as sensing, information processing, and optical coherence tomography.

## Results

**Design of 3-dB couplers with topological valley edge modes**

Consider a conventional 3-dB coupler, such as DC, consists of four ports [labeled with ①-④ in Fig. 1a]. Its performance is described by a 4×4 scattering matrix $\mathbf{S} = (s_{mn})_{4\times 4}$, where $s_{mn}$ represents the field intensity ratio between the output port $n$ and the input port $m$. At the wavelength of $\lambda_1$, the ideal 3-dB power splitting occurs, dividing the input signal from port 1 into two signals of equal powers at the port 3 and port 4, i.e., $|s_{31}|^2 = |s_{41}|^2$. However, this 3-dB splitting is achieved at a single wavelength due to the interference between waveguide modes. Any fabrication dimensional errors or changes in working temperature would degrade the 3-dB splitting at the designed wavelength, significantly hindering the device performance. To overcome these limitations and achieve a high-performance 3-dB coupler with wavelength-independent splitting ratio over a wide bandwidth, we incorporate topological principles into the design of nanophotonic devices. As illustrated in Fig. 1b, we have designed a topological 3-dB coupler based on the VPC slab, implemented on the silicon-on-insulator platform. Sandwiched between the silica substrate and the cladding, the VPC slab is constructed by a honeycomb lattice of equilateral triangular holes with a lattice constant of $a$ = 440 nm in a 220-nm-thick silicon layer. To create a nontrivial band gap capable of supporting transverse-electric-like edge modes, we set the side lengths of two holes, denoted as $s_1$ and $s_2$, to different values (Fig. 1c). We choose the parameters of VPCs according to the photonic band structures [see details in Supplementary Section G]. For VPC1, we set $s_1$ = 0.4$a$ and $s_2$ = 0.7$a$, while for VPC2, we set $s_1$ = 0.7$a$ and $s_2$ = 0.4$a$. It results in a common band gap spanning from 1486 nm to 1585 nm for both VPC1 and VPC2 [see details in Supplementary Section A]. Note that these two VPCs have distinct topological invariants, which are characterized by the valley Chern numbers with opposite signs, i.e. $C_v$ < 0 for VPC1 and $C_v$ > 0 for VPC2. Due to the bulk-edge correspondence, the domain wall between VPC1 and VPC2 (Fig. 1c) can support valley-dependent edge modes, whose propagation direction sgn[$v_g$] (the sign of group velocity) is equivalent to $\tau_z$ sgn[$C_{vb}$]. Here $C_{vb}$ is the valley Chern number of VPC at the bottom domain wall and $\tau_z$ = 1 ($\tau_z$ = -1) for K valley (K' valley). Edge21 supports rightward-propagating K valley and leftward-propagating K' valley edge modes, while Edge12 supports rightward-propagating

K' valley and leftward-propagating K valley edge modes. This valley contrasting edge modes suggests the suppression of the scattering between two valley edge modes.

Utilizing the two distinct edges, we design a topological coupler capable of achieving ideal 3-dB splitting. The device consists of a harpoon-shaped structure formed by connecting two different edges, resulting in four ports (Fig. 1b). To analyze its performance, we consider four cases when the light is incident from different ports. For example, when the light is incident from port 1, the rightward K valley edge mode along Edge21 is excited. Because of the inter-valley scattering suppression, both the leftward K' valley edge mode along Edge21 (backward to port 1) and the rightward K' valley edge mode along Edge12 (forward to port 2) are forbidden, indicated by two red crossings in Fig. 1b. This leads to $s_{11} = s_{21} = 0$, and consequently the incident light can only propagate to port 3 or port 4. As the coupler has the mirror symmetry, the incident light should be equally divided into port 3 and port 4, i.e., $s_{31} = s_{41} = \sqrt{2}/2$. This is confirmed by the full wave simulation in Fig. 1d. When light is incident from port 1, the simulated $H_z$ field shows that the incident waves is equally divided and propagates along both the upper and lower channels, resulting the 3-dB splitting ratio between port 3 and port 4. Likewise, we determine the other elements of the scattering matrix and ultimately obtain the scattering matrix of the proposed topological coupler:

$$S = \frac{\sqrt{2}}{2} \begin{pmatrix} 0 & 0 & 1 & 1 \\ 0 & 0 & 1 & -1 \\ 1 & 1 & 0 & 0 \\ 1 & -1 & 0 & 0 \end{pmatrix} \quad (1)$$

The derivation of each element of the scattering matrix is detailed in Supplementary Section C.

We note that the scattering matrix described above represents the ideal 3-dB splitting over a broad bandwidth that determined by the operation range of valley edge modes. The scattering matrix is derived from topology and symmetry, which are independent to wavelength and are always valid in the range of topological edge modes in band gap. Since no additional wavelength condition is required, the matrix elements are wavelength-independent in this range. Therefore, the 3-dB splitting can be achieved in the operation range of edge modes. This remarkable characteristic is protected by the nontrivial valley-Hall topology and the mirror symmetry of the structure.

**Experimental characterization of splitting ratio and robustness of topological 3-dB coupler**

In this section, we present the experimental investigation of the topological 3-dB coupler. The topological couplers are fabricated on the SOI platform using a top-down nanofabrication process [see Methods for fabrication details]. The optical microscope image of the whole sample, including topological coupler, strip waveguides and grating couplers, is shown in Fig. 2a. For the convenience of measurement, the grating couplers of port 1 and port 2 are placed on the left side, and the couplers of port 3 and port 4 are placed on the right side. Figure 2b gives the scanning electron microscope (SEM) images of the splitting junction formed by VPC1 and the VPC2. The junction is constructed by four edges, of which three of them are Edge21 and one is Edge12.

To confirm the broadband and robust 3-dB splitting performance in experiment, a fiber-to-waveguide alignment system is set up to image the light energy output from the grating couplers and to measure the transmission spectra [see Methods for the optical measurement setup]. Figures 2c and 2d show the measured splitting ratio spectra for light input from four different ports. The splitting ratio in Fig. 2c is defined as the ratio of transmittances between port 3 and port 4 when the light is incident from port 1 (red line) or port 2 (blue line). The incident waves are equally divided at the junction and propagate along the upper and lower channels to the output ports, resulting in a near-3-dB splitting ratio between port 3 and port 4. Similarly, the splitting ratio of the transmittances between port 1 and port 2 is depicted in Fig. 2d, when the light is incident from port 3 (red line) or port 4 (blue line). The splitting ratio spectra remain flat and close to 3-dB within the topological band gap (from 1530 nm to 1583 nm) for all four cases, which is in good agreement with the simulation results [see details in Supplementary Section D]. Note that the experimental results are difficult to reach ideal 3-dB ratio, as the fabricated samples may have minor deviations from rigorous mirror symmetry and difference in coupling efficiency between two output ports. We also show the far-field microscope images in Figs. 2e and 2f, with the light incident from port 1 and port 3 at a wavelength of 1550 nm, respectively. After splitting in the 3-dB coupler, the equally divided light is radiated into free space by grating couplers. In Fig. 2e, we observe strong and symmetric radiation patterns between the gratings of port 3 and port 4, while that of port 2 is comparatively low due to inter-valley-scattering suppression. Similar results are observed in Fig. 2f when the light incident from port 3. However, the radiation patterns from port 1 and port 2 exhibit slight difference, primarily stemming from the varying insertion losses of port 1 and port 2.

During the processes of nanofabrication, fabrication errors are inevitable, including dimensional error, sidewall roughness, defects and dislocations. Dimensional error is common for the fabrication of topological 3-dB coupler, corresponding to variations of $\pm 10$ nm in the size of the VPC holes. Consequently, we demonstrate the tolerance analysis of topological 3-dB couplers against dimensional errors, which has practical significance for on-chip nanodevices. As shown in Fig. 3a, the side lengths of triangular holes are varied by $\Delta s$. Far-field microscope images show that the radiated patterns at $\lambda$ = 1550 nm maintain symmetric profiles between the gratings of port 3 and port 4, even with $\pm 10$ nm variations in the size of the VPC holes (Fig. 3a). Due to broadband operation of near 3-dB in topological coupler, the splitting ratio nearly remain as 3-dB in a wavelength range even the spectra shift. Originated from the broadband property, the experimental ratio spectra of these three cases (Fig. 3b) keep in a flat platform ranging from 1533 nm to 1581 nm (orange region), albeit with a blue shift as $\Delta s$ increases. For comparison, we also consider the dimensional error of conventional DC, i.e. the gap width $\Delta g$ between two coupled waveguides [see schematics in Fig. 3c]. For the cases of $\Delta g = \pm 10$ nm, the far-field microscope images have asymmetric radiated patterns between port 3 and port 4, although the case of $\Delta g = 0$ is symmetric. In contrast to the near 3-dB flat platform in topological coupler, the splitting ratio spectra of DC are oblique slopes, as shown in Fig. 3d. The 3-dB splitting of DC only occurs at a single wavelength which has a red shift as $\Delta g$ increases. Consequently, the performance of conventional DC is strongly sensitive to the size variations, and this is unwanted when nanodevices with precise splitting ratio are required. Further to cascading several DCs, the sensitivity to wavelength and dimensional error will largely increase. In comparison, topological couplers still have flat splitting ratio even though they are cascaded under consideration of dimensional errors. It implicates topological coupler will exhibit better robustness in the cascaded devices of PICs, such as optical neural networks and quantum logic gates. Even though we introduce non-uniform dimensional errors which break the mirror symmetry, the splitting ratio still exhibits robustness as the structure of coupler and the dispersion of edge modes are not seriously changed. In addition to fabrication errors, temperature drift will variate the refractive index of semiconductor materials and thus affect the performance of 3-dB couplers. We show that topological couplers also outperform conventional couplers for the robustness against temperature changes. [see details in Supplementary Section D].

**On-chip interferometer based on topological coupler**

In addition to the demonstration of the principles of topological physics, another important goal in topological nanophotonics is to achieve robust optical functionality even in the presence of fabrication errors. To verify this capability, we utilize the topological 3-dB coupler to construct an on-chip Michelson-like interferometer. The schematic in Fig. 4a illustrates the interferometer's configuration, which consists of a 3-dB coupler and a pair of distributed Bragg reflectors (DBRs) connected to the output waveguides of port 3 and port 4. The DBRs have reflectivity of over 90% from 1500 nm to 1600 nm. The magnified SEM image of the purple box in Fig. 4a shows the coupling area of port 1 (same to port 3 and port 4), where a 2-μm-width strip waveguide is connected to the VPC. The first 18 rows of VPC holes are modified to introduce adiabatic-like mode evolution, which would improve the coupling efficiency. For port 2 (see magenta box in Fig. 4a), a line-defect photonic crystal waveguide is introduced and connected to a 400-nm-wide strip waveguide.

This interferometer design can be used to carry out distance measurements. For example, in Fig. 4 the two output waveguides have a length difference of 100 μm. The shorter one serves as the reference arm, while the longer one the measuring arm. In the experiment, the light is input from port 1, split by the coupler, and directed to port 3 and port 4 [see red arrows]. The two coherent light beams are then reflected by DBRs, injected into port 3 and port 4, and interfere at the splitting junction of the coupler [see blue arrows]. The difference in waveguide lengths between the reference and measuring arms introduces a phase difference $\Delta\phi$, which determines the transmittance to port 2:

$$T_2(\lambda) = \frac{1}{2}\left[1 - \cos\Delta\phi(\lambda)\right]. \tag{2}$$

Owing to the wavelength-dependent nature of the waveguide mode dispersion (i.e., $\Delta\phi$ is a function of wavelength), the transmission spectrum of port 2 exhibits periodic dips (Fig. 4b). To evaluate the performance of the interferometer, we consider a critical parameter called the extinction ratio (ER), defined as the ratio of the transmittance at the dip wavelength to that at the nearest peak wavelength.

Figure 4c presents the experimental results of the ER as a function of wavelength. The ER is larger than 20 dB in the wavelength range of 1530 nm to 1583 nm (the region shaded in orange). This indicates that the interference functionality can cover a bandwidth of 53 nm with a 20-dB ER.

Furthermore, the length difference between two waveguides can be extracted by applying Fourier transform on the transmission spectrum. After introducing the dispersion of strip waveguide, we can obtain the length difference from the peak position of Fourier transform result [see details in Supplementary Section H]. In Fig. 4d, the Fourier transformed magnitude shows a peak around 100 μm (green line), which agrees reasonably with the practical sample parameter. Notably, even in the presence of the dimensional error of $\Delta s = \pm 10$ nm, the extracted peaks remain around 100 μm (red and blue lines in Fig. 4d), confirming the robustness of our interferometer against dimensional errors. It is important to point out that the topological interferometer offers the potential for extensive functionalities. Furthermore, based on the same principle, we can also measure the distance in free space. By replacing one of the DBRs with a sample containing multiple reflection interfaces, we can reconstruct the axial image of a sample with different layers. This opens up new ways for applications in coherent imaging and analysis of complex media. Moreover, the topological interferometer enables not only the measurement of the length difference between two arms but also the sensing of physical quantities that change the refractive index, such as temperature, pressure, and material, etc. Leveraging the high extinction ratio provided by the topological coupler, our design holds promise for high-performance on-chip optical coherent tomography and temperature/pressure sensors.

## Discussion

We have designed and realized a topological coupler that maintains a 3-dB splitting ratio over a broad wavelength range, even in the presence of common dimensional errors during practical fabrication processes. The underlying principle of the 3-dB splitting origins from the valley-Hall topology and mirror symmetry, which we have theoretically elucidated by deriving a wavelength-insensitive scattering matrix. The tolerance analysis confirmed the robustness of the topological 3-dB coupler against dimensional errors, as compared to the conventional directional couplers (DCs). Furthermore, the proposed topological coupler was utilized to construct an on-chip interferometer capable of extracting the length difference between two arms. The interferometer exhibits remarkable resilience to dimensional errors, which is crucial for practical applications. This finding gives a systematic analysis on the advantages offered by topologically-protected edge modes and explicitly demonstrates the crucial performance of photonic structures. The proposal of topological splitting structure paves the way for the development of practical on-chip nanophotonic devices with topological protection. For example, the length difference measurement function can be applied in optical coherence tomographs and frequency modulated continuous wave LiDARs. To do this, the insertion loss between photonic crystals and strip waveguides should be further reduced to improve the signal-noise ratio of nanophotonic topological devices. The inverse design is anticipated to be applied for improving the performance of topological devices. In addition, the bandwidth and the footprint of topological coupler can be optimized based on the properties of topological valley edge modes [see details in Supplementary Sections G and I]. Based on VPCs, various kinds of on-chip photonic devices have been realized, such as asymmetric splitters [52], photonic switches [53] and add-drop filters [54]. The design method based on topological photonic principle is demonstrated to be generalizable to many kinds of on-chip devices. In consideration of the generalizability of topological-photonic-design method, the designs of topological 3-dB coupler and interferometer show promising applications for on-chip sensors, modulators, photodetectors, and wavelength division (de)multiplexers.

## Materials and methods

**Numerical simulation.**

The transmission spectra and field distributions in our devices were simulated using Ansys Lumerical FDTD software, utilizing the finite-difference time-domain (FDTD) method. The band structures and eigen modes of VPCs were calculated using the eigenmode solver of MIT Photonic Bands (MPB) [55]. To ensure the accurate evaluation, 3D full-wave simulations were conducted with the same structural parameters as those used in the experiment. For the propagation simulations, perfectly matched layers (PMLs) were used at the boundaries of the simulation domains to absorb the output light and the stray light. In contrast, the band structure calculations employed periodic boundary conditions to account for the periodicity of the photonic crystal structure.

**Sample fabrication.**

We fabricated the experimental sample on a silicon-on-insulator (SOI) wafer, with a 220-nm-thick silicon top layer and a 3-μm-thick buried oxide layer. The pattern of the devices was defined using electron-beam lithography (Vistec EBPG 5200+). Inductively coupled plasma etching technique (SPTS DRIE-I) was employed to transfer the pattern onto the silicon layer. The strip waveguides, photonic crystals, and distributed Bragg reflectors (DBRs) were etched to a depth of 220 nm, while the grating couplers was etched to a depth of 70 nm. Finally, a silica upper-cladding layer was deposited on top of the silicon layer.

**Optical characterization.**

For the measurements of transmission spectra and far-field images, three tunable continuous-wave lasers (Santec TSL-550/710) were utilized to cover the wavelength range from 1260 nm to 1640 nm. The incident light was launched into a fiber and then coupled to the input waveguide using a grating coupler. After passing through the devices based on VPCs, the output light was either coupled into a fiber or emitted into free space via grating couplers. To get the transmission spectra, the output light coupled to the fiber was detected by an optical power meter (Santec MPM-210). For obtaining far-field images, the output light emitted into free space was collected by a varifocal microscope objective and then imaged using an InGaAs CCD camera (Xenics Bobcat-640-GigE).

## Data availability

The authors declare that all data supporting the findings of this study are available within the paper and its Supplementary Information files.

## Acknowledgements

This work was supported by National Key Research and Development Program of China (Grant No. 2022YFA1404304), National Natural Science Foundation of China (Grant Nos. 62035016, 12274475, 12074443, 62105200), Guangdong Basic and Applied Basic Research Foundation (Grant Nos. 2023B1515040023, 2023B1515020072), Fundamental Research Funds for the Central Universities, Sun Yat-sen University (23lgbj021, 23ptpy01).

## Conflict of interest

All authors agree with the submission and declare that they have no conflict of interest.

**Figures and Captions**

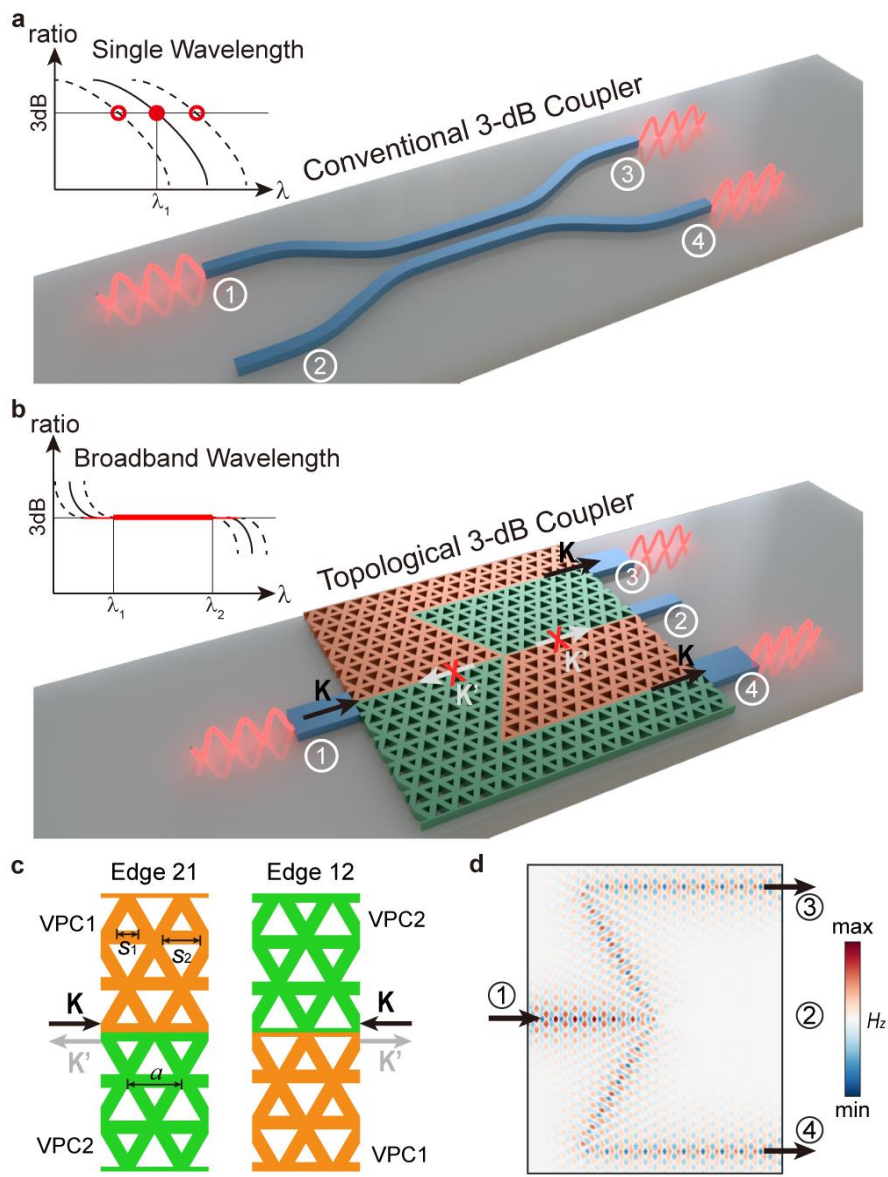

**Figure 1 | Conventional and topological 3-dB couplers. a, b,** Beam splitting in (**a**) a conventional 3-dB directional coupler and (**b**) a topological 3-dB coupler composed of two different VPCs. The numbers in circles label four different ports. Light input from port 1 is divided into ports 3 and 4. The inset shows the splitting ratio as a function of wavelength, in which the dash lines represent the cases with dimensional errors. The 3dB splitting is achieved at only one single wavelength $\lambda_1$ for the conventional coupler. In contrast, the topological coupler achieves the ideal 3-dB splitting in a broadband wavelength even when dimensional errors are introduced. The white arrows with crosses indicate that no backscattering to port 1 and no straight transmission to port 2 are allowed. **c**, Structures of Edge21 and Edge12 which are constructed with VPC1 and VPC2. Structural parameters such as $s_1$, $s_2$ and $a$ are marked. The black (gray) arrows denote the propagation directions of edge modes at K (K') valleys, respectively. **d**, $H_z$ field within the topological 3-dB coupler when the light is incident from port 1. No backscattering and no through transmission are observed due to the inter-valley scattering suppression. The incident light is divided equally into two output ports, resulting in the 3-dB splitting.

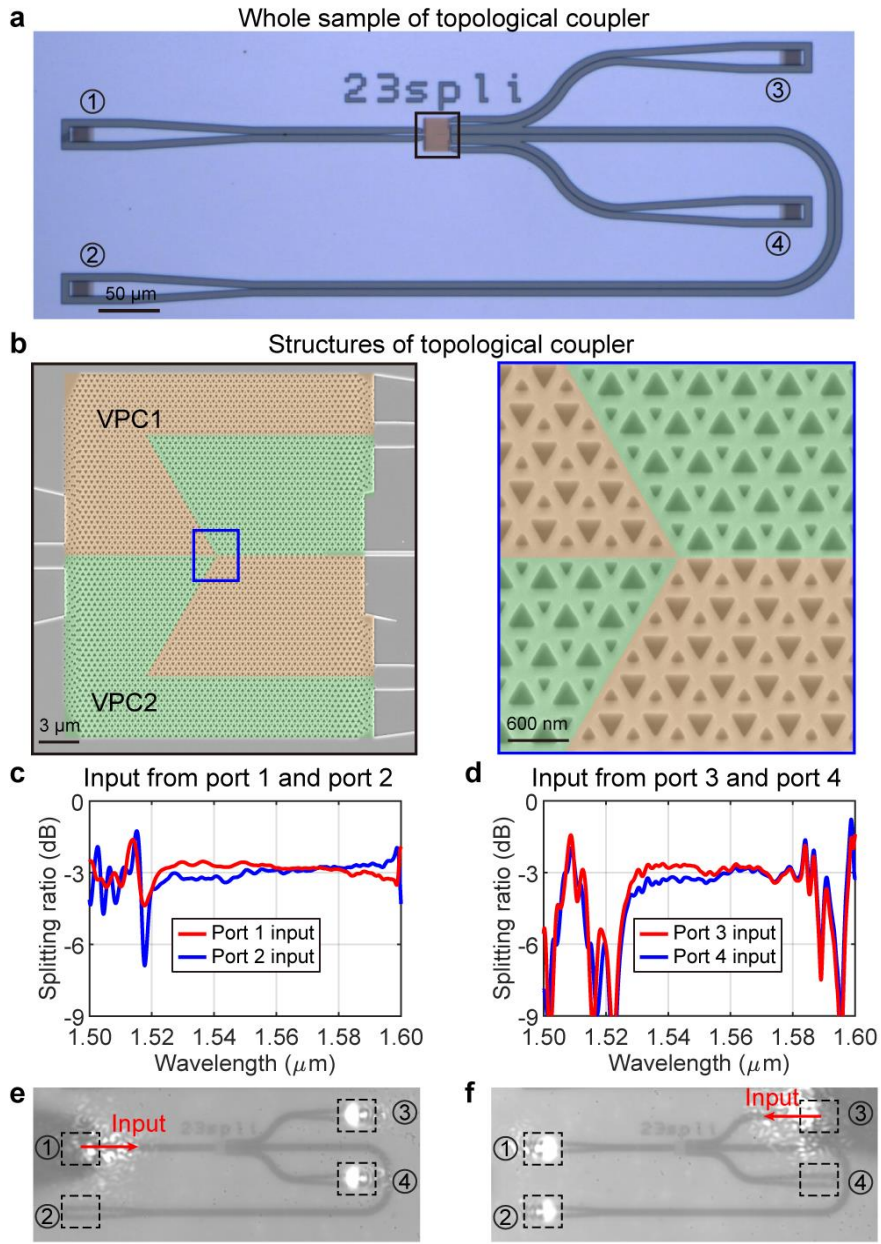

**Figure 2 | Broadband 3-dB splitting of the topological coupler. a**, Optical micrograph of the fabricated topological 3-dB coupler. **b**, Scanning electron microscope (SEM) images of the splitting junction between two topologically distinct VPCs (i.e. VPC1 and VPC2). **c, d**, Experimental splitting ratio spectra of the fabricated topological 3-dB coupler when light is incident from different ports. **e, f**, Far-field images taken from an optical microscope system for inputting light of 1550 nm from (**e**) port 1 and (**f**) port 3. The dashed boxes and numbers 1-4 in circles denote the grating couplers for four ports. The red arrows denote the input ports.

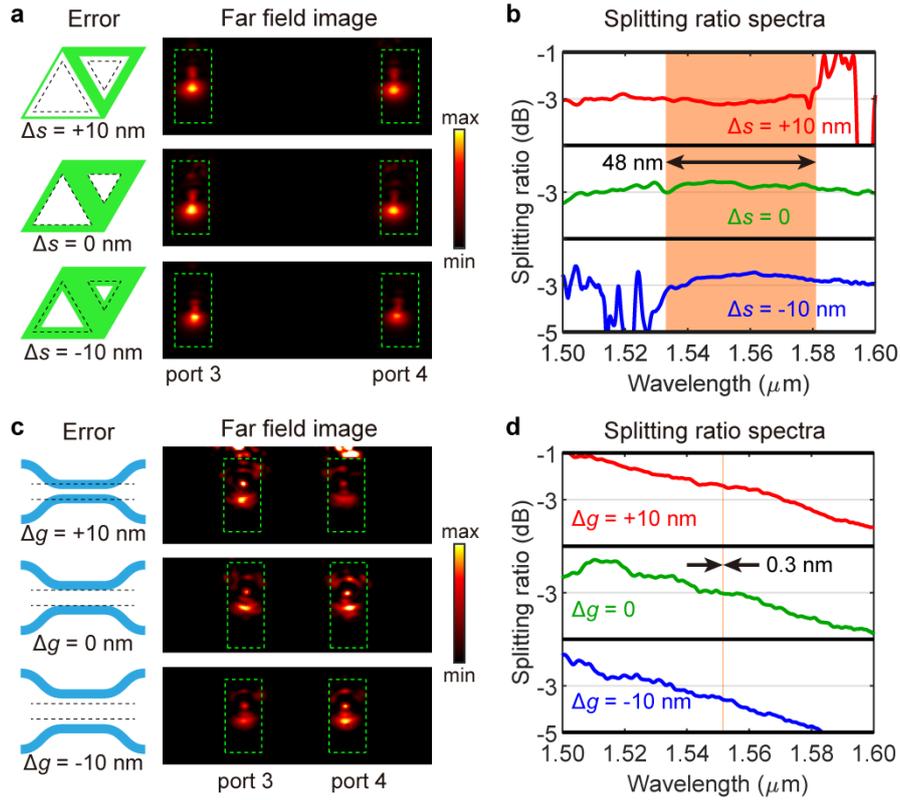

**Figure 3 | Tolerance analysis of the topological and conventional 3-dB couplers. a** and **c**, Schematic of the dimensional errors and the corresponding radiation patterns of far-field microscope images in (**a**) topological couplers and (**c**) conventional directional couplers. The patterns are collected from the grating couplers of port 3 and port 4, when light of 1550 nm is input from port 1. The green dashed boxes denote the grating couplers. **b** and **d**, Measured splitting ratio spectra for (**b**) topological couplers and (**d**) conventional couplers with different dimensional errors when inputting light from port 1. The orange regions in **b** and **d** represent the range where that splitting ratio for all three cases keeps within $(3 \pm 0.6)$ dB. The region spans from 1533 nm to 1581 nm in **b** and only 0.3 nm (from 1551.43 nm to 1551.73 nm) in **d**.

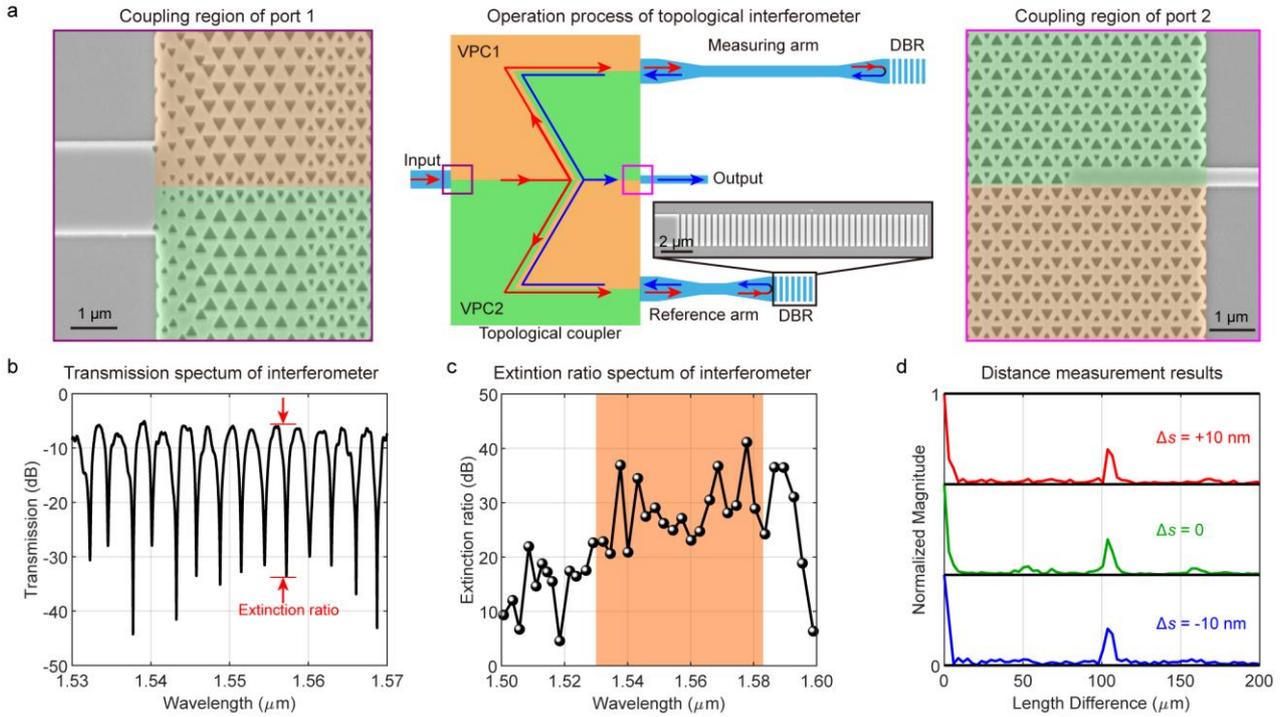

**Figure 4 | Distance measurement with topological interferometers. a**, Structure and operation principle of the topological interferometer. Middle: Illustration of the topological interferometer constructed by the 3-dB topological coupler and a pair of DBRs that placed at the end of the reference and measuring arms. The red arrows represent the forward propagating waves, while the blue arrows represent those reflected by the DBRs. The inset shows the SEM image of DBR. Left: SEM image of the coupling region of port 1. Right: SEM image of the coupling region of port 2. The coupling regions are designed to reduce the insertion loss of photonic crystal waveguides. **b**, Transmission spectrum of topological interferometer in the C band. The red arrows show the extinction ratio. **c**, Extinction ratio spectra of topological interferometer. The operation range from 1530 nm to 1583 nm is colored in orange, where the extinction ratio is larger than 20 dB. **d**, Distance measurement with dimensional errors. The magnitude for certain length difference represents the intensity of light beams experiencing corresponding length difference in two arms. The magnitude is normalized by the zero-length-difference value, which represents the average of transmission spectrum.